\documentclass[aps,prd,showpacs,floats]{revtex4}
\usepackage{epsfig}
\usepackage{dcolumn}
\usepackage{bm}
\usepackage{amsmath}
\usepackage{amsfonts}
\usepackage{amssymb}
\usepackage{graphicx}
\newcommand{\bea}{\begin{eqnarray}}
\newcommand{\eea}{\end{eqnarray}}
\newcommand{\bwt}{\begin{widetext}}
\newcommand{\ewt}{\end{widetext}}
\def\be{\begin{equation}}
\def\ee{\end{equation}}
\begin{document}
\title{Scalar Mesons and glueballs in $Dp-Dq$ hard-wall models}
\author{Chao Wang$^{1}$}
\author{Song He$^{1}$}
\author{Mei Huang$^{1,2}$}
\author{Qi-Shu Yan$^{3}$}
\author{Yi Yang$^{4,5}$}
\affiliation{$^{1}$ Institute of High Energy Physics, Chinese Academy of Sciences,
Beijing, China}
\affiliation{$^{2}$ Theoretical Physics Center for Science Facilities, Chinese Academy of
Sciences, Beijing, China}
\affiliation{$^{3}$ Department of Physics, University of Toronto, Toronto, Canada}
\affiliation{$^{4}$ Department of Electrophysics, National Chiao-Tung University,
Hsinchu, Taiwan}
\affiliation{$^{5}$ Physics Division, National Center for Theoretical Sciences, Hsinchu,
Taiwan}

\date{\today }

\begin{abstract}
We investigate light scalar mesons and glueballs in the $Dp-Dq$ hard-wall models,
including $D3-Dq$, $D4-Dq$, and $D6-Dq$ systems. It is found that only
in the $D4-D6$ and $D4-D8$ hard wall models, the predicted masses of
the ${\bar q} q$ scalar meson $f_0$, scalar glueball are consistent 
with their experimental or lattice results. This indicates that $D4-D6$ and 
$D4-D8$ hard-wall models are favorite candidates of the realistic 
\textit{holographic} QCD model.
\end{abstract}

\pacs{11.25.Tq, 11.10.Kk, 11.25.Wx, 12.38.Cy}
\maketitle

\section{Introduction}

In recent years, there have been intense studies on scalar mesons and 
scalar glueballs and their mixing, \textit{e.g.} see 
Refs. \cite{Scalar,Scalar-lightGB,Glueball-Review} 
and references therein. 

The glueball spectrum has attracted much attention more than three decades
\cite{Glueball-first}.  Study particles like
glueballs where the gauge field plays a more important dynamical
role than in the standard hadrons, offers a good opportunity of
understanding the nonperturbative aspects of QCD. 
The complexity of determing the glueball states lies in that gluonic
bound states always mix with ${\bar q} q$ states. For example, one has
to distinguish the lightest scalar glueball state among other scalar mesons
observed in the energy range below $2 {\rm GeV}$. Though the pseudoscalar, 
vector and axial-vector, and tensor mesons with light quarks have been 
reasonably well known in terms of their $SU(3)$ classification
and quark content, the scalar meson sector, on the other hand, is much
less understood in this regard. There are 19 states which are more than
twice the usual ${\bar q}q$ nonet as in other sectors.

Despite of extensive study from both experimental side and theoretical side, no 
conclusive answer has been obtained on scalar mesons and scalar glueballs. 
One possible scenario is: The lightest scalars $\sigma, \kappa, f_0, a_0$ below 
$1 {\rm GeV}$ 
make a full $SU(3)$ flavor nonet. The inversion of the $\kappa$ and $f_0$ or $a_0$ 
mass ordering, suggests that these mesons are not naive ${\bar q}q$ states, one 
natural explanation
for this inverted mass spectrum is that these mesons are diquark and antidiquark
bound states, or tetraquark states \cite{Jaffe-tetraquark}.  Above $1 {\rm GeV}$, the
nonet ${\bar q}q$ mesons are made of an octet with largely unbroken $SU(3)$
symmetry and a fairly good singlet which is $f_0(1370)$. The other left scalar
meson $f_0(1710)$ is identified as an almost pure scalar glueball with a
$\sim 10 \%$ mixture of ${\bar q}q$, which is  supported from lattice calculation
\cite{glueball-lattice} and experimental observation of the copious $f_0(1710)$
production in radiative $J/\psi$ decays \cite{pdg06}. 

Recently, the discovery of the gravity/gauge duality, or anti-de Sitter/conformal
field theory (AdS/CFT) correspondence \cite{dual,m5} provides a revolutionary
method to tackle the problem of strongly coupled gauge theories, for reviews
see Ref. \cite{Reviews}. Many efforts have been invested in examining meson
spectra, baryon spectra, see \textit{e.g.} Refs.
\cite{Bottomup,baryon}, as well as in the
glueball sector \cite{glueball-topdown,glueball-bottomup}.
It is widely expected that this new analytical approach can shed some light on 
our understanding of the nonperturbative aspects of QCD.

The string description of realistic QCD has not been successfully formulated yet.
By using AdS/CFT correspondence to study non-conformal
field theory like QCD, the usual way of breaking conformal symmetry is by introducing
a hard infrared (IR) cut-off, i.e. the hard-wall $AdS_5$ model or introducing a smooth
cut-off through a dilaton background field, i.e. the soft-wall $AdS_5$ model.
One can extend the AdS/CFT correspondence to a more
general case, and expect the realistic QCD is dual to a non-conformal $Dp$ brane system,
like the $D4-D8/{\bar D8}$ system, i.e. the Sakai-Sugimoto model \cite{SS}.
In Ref. \cite{HHYY},  we have investigated the general embedding $Dp-Dq$ systems, where
the $N_{c}$ background $Dp$-brane describes the effects of pure QCD theory, while the 
$N_{f}$ probe $Dq$-brane is to accommodate the fundamental flavors. 

The motivation of this paper is to investigate the scalar meson and glueball 
spectra in the general embedding $Dp-Dq$ systems, and study which $Dp-Dq$ system 
is more close to the dual theory of realistic QCD. Our finding is that in the $D4-D6$ and 
$D4-D8$ hard wall models, the predicted masses of the ${\bar q} q$ scalar meson $f_0$ 
and the scalar glueball are consistent with  their experimental or lattice results, which indicates 
that $D4-D6$ and $D4-D8$ hard-wall models are favorite candidates of the realistic 
\textit{holographic} QCD model.
Because this paper is an attempt to describe light mesons and glueballs in one 
\textit{holographic} model, we will leave the mixing
between scalar mesons, tetraquark states and glueballs for future studies. 

The paper is organized as following: After the introduction, we briefly introduce 5-dimension metric
structure of the $Dp-Dq$ system in type II superstring theory in Sec. \ref{sec-DpDq}.
Then in Sec. \ref{sec-spectra}, we give the equation of motion for mesons and glueballs, and
we investigate the meson spectra and glueball spectra. At the
end we give discussions and conclusions in Sec.\ref{sec-summary}.

\section{The $D_p-D_q$ system}

\label{sec-DpDq}

We have investigated the $Dp-Dq$ systems in Ref. \cite{HHYY}, however, in order to keep
this paper self-contained,  in the following, we give a brief introduce on
the $Dp-Dq$ branes system in type II superstring theory. In the $Dp-Dq$ system, the
$N_{c}$ background $Dp$-brane describes the effects of pure gauge theory, while
the $N_{f}$ probe $Dq$-brane is to accommodate the fundamental flavors which
has been introduced by Karch and Katz \cite{Karch:2002sh}.

The near horizon solution of the $N_{c}$\ background $Dp$-branes in type II
superstring theory in 10-dimension space-time is \cite{Dp-brane}%
\begin{equation}
ds^{2}=h^{-\frac{1}{2}}\eta _{\alpha \beta }dx^{\alpha }dx^{\beta }+h^{\frac{1}{2}%
}\left( du^{2}+u^{2}d\Omega _{8-p}^{2}\right) ,  \label{metric}
\end{equation}%
where $\alpha,\beta=0,\cdots,p$, $\eta _{\alpha \beta }={\rm
diag}(-1,1,1,...)$, and the warp factor $h\left( u\right) =\left(
R/u\right) ^{7-p}$ and $R$ is a constant
$R=\left[ 2^{5-p}\pi ^{\left( 5-p\right) /2}\Gamma \left( \frac{7-p}{2}%
\right) g_{s}N_{c}l_{s}^{7-p}\right] ^{\frac{1}{7-p}}$.
The dilaton field in this background has the form of
$e^{\Phi }=g_{s}~h\left( u\right) ^{\frac{\left( p-3\right) }{4}}$.
The effective coupling of the Yang-Mills theory is  $g_{eff}\sim g_{s}N_{c}u^{p-3}$,
which is $u$ dependent. This $u$ dependence corresponds to the RG flow in
the Yang-Mills theory, i.e. the effective $g_{eff}$\ coupling constant
depends on the energy scale $u$. In the case of $D3$-brane, $g_{eff}\sim
g_{s}N_{c}$\ becomes a constant and the dual Yang-Mills theory is $\mathcal{N%
}=4$ SYM theory which is a conformal field theory. The curvature\ of the
background (\ref{metric}) is $\mathcal{R}\sim \frac{1}{l_{s}^{2}g_{eff}}$,
which reflects the string/gauge duality - the string on a background of
curvature $\mathcal{R}$\ is dual to a gauge theory with the effective
coupling $g_{eff}$. To make the perturbation valid in the string side, we
require that the curvature is small $\mathcal{R}\ll 1$, which means that the
effective coupling in the dual gauge theory is large $g_{eff}\gg 1/l_{s}^{2}$%
. In the case of $D3$-brane, the curvature $\mathcal{R}$\ becomes a
constant, and the background (\ref{metric}) reduces to a constant curvature
spacetime - $AdS_{5}\times S^{5}$.

The coordinates transformation (for the cases of $p\neq 5$) 
$u=\left( \frac{5-p}{2}\right) ^{\frac{2}{p-5}}R^{\frac{p-7}{p-5}}z^{\frac{2}{%
p-5}}$,
brings the above solution (\ref{metric}) to the following \emph{Poincar\'{e}
form,}%
\begin{equation}
ds^{2}=e^{2A\left( z\right) }\left[ \eta _{\alpha \beta
}dx^{\alpha }dx^{\beta }+dz^{2}+\frac{\left( p-5\right)
^{2}}{4}z^{2}d\Omega _{8-p}^{2}\right] .
\end{equation}%

We then consider $N_{f}$ probe $Dq$-branes with $q-4$ of their
dimensions in the $S^{q-4}$ part of $S^{8-p}$, with the other
dimensions in $z$ and $x^{\alpha }$ directions. The induced $q+1$
dimensions metric on the probe branes is given as
\begin{equation}
ds^{2}=e^{2A(z)}\left[ \eta _{\mu \nu }dx^{\mu }dx^{\nu }+dz^{2}+\frac{z^{2}}{%
z_{0}^{2}}d\Omega _{q-4}^{2}\right] ,  \label{induced}
\end{equation}
where $\mu,\nu=0,\cdots,3$, $\eta _{\mu \nu }={\rm
diag}(-1,1,1,1)$, and the metric function of the warp factor only
includes the logarithmic term
\begin{align}
A(z)=-a_{0} ~\mathrm{ln} z, ~~with ~~a_{0}=\frac{p-7}{2\left( p-5\right) },
\end{align}
and the dilaton field part takes the form of
$e^{\Phi(z)} = \ g_{s}\left( \frac{2}{5-p}\frac{R}{z}\right) ^{\frac{\left(
p-3\right) \left( p-7\right) }{2\left( p-5\right) }}$,
which gives
\begin{equation}
\Phi(z) \sim d_{0} \ln z, ~with ~d_{0}=-\frac{\left( p-3\right) \left(
p-7\right) }{2\left( p-5\right) }.  \label{Phi}
\end{equation}
The metric (\ref{induced}) is conformal to $AdS_5\times S^{q-4}$.

A 5-dimension (5D) scalar field $X(x,z)$ can be described by  the  action in the
gravitational background as
\begin{equation}
I_{S=0}=\frac{1}{2}\int d^5x\,\sqrt{g}\,e^{-\Phi(z)}\,\,
\left[ \partial_{N}X \partial^{N}X + m_{5,X}^2 X^2 \right]\;,
\,\,\,\,\label{actionscalmass}
\end{equation}
For higher spin fields, we can have the effective 5D
action described by tensor fields as
\begin{align}
I_{S>0} & =\frac{1}{2}\int d^{5}x\sqrt{g}\,\,e^{-\Phi(z)}\bigg \{%
\Delta_{N}\phi_{M_{1}\cdots M_{S}}\Delta^{N}\phi^{M_{1}\cdots M_{S}}  \notag
\\
& +m_{5,\phi}^{2}\phi_{M_{1}\cdots M_{S}}\phi^{M_{1}\cdots M_{S}}\bigg \},
\label{spin}
\end{align}
where $\phi_{M_{1}\cdots M_{S}}$ is the tensor field and $M_i$ is the tensor
index. The value of $S$ is equal to the spin of the field.
The parameters $g$ and $\Phi(z)$ are the induced $q+1$ dimension metric
and dilaton field as shown in Eq. (\ref{induced}) and (\ref{Phi}).
$m_{5,X}^2$ and $m_{5,\phi}^2$ are the 5D mass square of the bulk fields.

By assuming that the gauge fields are independent of the internal space
$S^{q-4}$, after integrating out $S^{q-4}$, up to the quadratic terms and
following the standard procedure of dimensional reduction, we can decompose
the bulk field into its 4D components $\phi^n(x)$ and their fifth profiles $\psi_n(z)$.
The equation of motion (EOM) of the fifth profile wavefunctions
$\psi_n(z)$ for the general spin field including $S=0$ and $S>1$ can be
derived as
\begin{eqnarray}
\partial_z^{2}\psi_{n} - \partial_z B \cdot \partial_z\psi_{n} +\left(
M_{n}^{2} -m_{5}^{2}e^{2A}\right) \psi_{n} =0\,,  \label{hispin}
\end{eqnarray}
where $M_{n}$ is the mass of the 4-dimension field $\phi^{n}(x)$, and
\begin{eqnarray}
B=\Phi-k^{\prime} k A= \Phi+ k^{\prime} c_0 \mathrm{ln}z
\end{eqnarray}
is the linear combination of the metric background function and the dilaton
field, with $k^{\prime}=3$ for scalar field, and $k^{\prime}=2S-1$ for higher
spin fields. For simplicity, we have defined
$c_0=k a_0= -\frac{\left( p-3\right) \left( q-5\right) +4}{2\left( p-5\right)}$.
The parameter $k$ is a parameter depending on the induced metric (\ref%
{induced}) of the $Dq$ brane. After integrating out $S^{q-4}$, $k$ is
determined as
$k=-\frac{\left( p-3\right) \left( q-5\right) +4}{p-7}$.
It is obviously that $k$ depends on both $p$ and $q$. 

The parameters $c_{0}, d_{0}$ and the curvature for any $Dp-Dq$ system are listed in
Table \ref{c0d0}. 
\begin{table}[th]
\begin{center}
$%
\begin{tabular}{|c|c|c|c|c|c|c|c|}
\hline
$p$ & \multicolumn{2}{|c|}{$3$} & \multicolumn{3}{|c|}{$4$} &
\multicolumn{2}{|c|}{$6$} \\ \hline
$q$ & \multicolumn{1}{|c|}{$5$} & $7$ & $4$ & $6$ & $8$ & $4$ & $6$ \\ \hline
$c_{0}$ & \multicolumn{2}{|c|}{$1$} & $3/2$ & $5/2$ & $7/2$ & $-1/2$
& $-7/2$ \\ \hline
$d_{0}$ &
\multicolumn{2}{|c|}{$0$} & \multicolumn{3}{|c|}{$-3/2$} &
\multicolumn{2}{|c|}{$3/2$} \\ \hline
$\mathcal{R}$ & \multicolumn{2}{|c}{$1/\sqrt{3}$} &
\multicolumn{3}{|c}{$z^{-2}/\sqrt{36\pi }$} & \multicolumn{2}{|c|}{$6\sqrt{2}%
z^{6}$} \\ \hline
\end{tabular}%
\ $%
\end{center}
\caption{Theoretical results for the $D{p}-D{q}$ system.}
\label{c0d0}
\end{table}
We notice that $d_{0}=0$ for $D{3}$ background branes, \textit{i.e.} dilaton
field is constant in AdS$_{5}$ space. However, the dilaton field in a
general $Dp-Dq$ system can have a $\ln z$ term contribution, e.g. in the $%
D4-D8$ system $d_{0}=-3/2$. We also want
to point out that for pure $Dp-Dq$ system, the curvature is proportional to
the inverse of the coupling strength $g_{eff}$. For $D{3}$ background
branes, the curvature is a constant. The curvature for $D{4}$ background
branes is small at IR, and large at UV, its dual gauge theory is strongly
coupled at IR and weakly coupled at UV, which is similar to QCD. However,
the curvature for $D{6}$ background branes is large at IR, and small at UV,
its dual gauge theory is weakly coulped at IR and strongly coupled at UV,
which is opposite to QCD.

\section{Meson spectra and glueball spectra in the $Dp-Dq$ hard-wall models }
\label{sec-spectra}

In the following, we are going to investigate the scalar mesons and glueballs in the 5D $Dp-Dq$
model defined in Sec. \ref{sec-DpDq}.  Because here we are only interested in the 
light excitations,  we will use hard-wall models of $Dp-Dq$ system, i.e.
we  choose a slice of the 5D $Dp-Dq$ metric in the region of $0<z\leq z_m$.
$z_m$ will be fixed in each $Dp-Dq$ model with the mass of vector meson $\rho(770)$.
We will use the scenario in the introduction as reference for the scalar mesons and glueballs:
the mass of ${\bar q}q$ scalar meson $f_0$ is in the range of $1370-1500 ~{\rm MeV}$  
\cite{pdg06}, and 
the mass of scalar glueball $G_0(0^{++})$ is around $1710~ {\rm MeV}$  
\cite{pdg06,glueball-lattice}, while  the nonet 
below $1{\rm GeV}$ are tetraquark states.
We will also study several tensor glueballs for reference,  the lattice result \cite{glueball-lattice} 
shows that  the masses for tensor glueball $G_2(2^{++})$
and $G_3(3^{++})$ are around $2400~ {\rm MeV}$ and $3600~ {\rm MeV}$, respectively.

The key ingredients of the AdS/CFT correspondence is that it establishes
a one-to-one correspondence between a certain class of local
operators  in the $4D$ ${\cal N}=4$ superconformal gauge theory and
supergravity fields representing the holographic correspondents in the $AdS_{5}\times
S^{5}$ bulk theory.   In the bottom-up approach, we can expect a more general
correspondence, i.e. each operator ${\cal O}(x)$ in the 4D field theory
corresponds to a field $\phi(x,z)$ in the 5D bulk theory. To investigate the
meson and glueball spectra, we consider the lowest dimension operators with the
corresponding quantum numbers and defined in the field theory living on the $4D$ 
boundary.
According to AdS/CFT correspondence, the
conformal dimension of a ($f$-form) operator on the boundary is
related to the $m_5^{2}$ of its dual field in the
 bulk as follows \cite{m5} :
\begin{equation}
m_5^2=(\Delta-f)(\Delta+f-4)\;.   \label{Eq-m5}
\end{equation}
For non-conformal $Dp$ branes, the induced metric (\ref{induced})
is still conformal to an AdS metric as we mentioned before. We
thus assume the above correspondence can be extended to any
$Dp-Dq$ system in 5-dimension. In Table \ref{m5}, we list the
correspondent fields for mesons and glueballs considered, and
their 5D mass square.

\begin{table}[th]
\begin{center}
$%
\begin{tabular}{|c|c|c|c|c|c|c|c|}
\hline
    & $4D: {\cal O}(x) $&  $\Delta$ & $f$ & $m_5^2$   \\ \hline
 $f_0$ & ${\bar q}q$ & $3$ & $0$ & $-3$   \\ \hline
 $\rho$ & ${\bar q}\gamma_{\mu} q$ & $3$ & $1$ & $0$  \\ \hline
$G_0 $ & $F^2$ & $4$ & $0$ & $0$   \\ \hline
$G_{2}$ & $FD_{\mu_1}D_{\mu_2}F$ & $6$ & $2$ & $16$    \\ \hline
$G_3$ & $FD_{\mu_1}D_{\mu_2}D_{\mu_3}F$ & $7$ & $3$ & $24$   \\ \hline
\end{tabular}%
\ $%
\end{center}
\caption{5D mass square of mesons and glueballs in $D_{p}-D_{q}$ system.}
\label{m5}
\end{table}

The equation of motion Eq. (\ref{hispin}) can be simplified as
\begin{eqnarray}
- \psi_n^{\prime \prime} + V(z) \psi_n = M_n^2 \psi_n \, , \label{EOM}
\end{eqnarray}
where $V(z)$ takes the form of
$V(z)= \frac{B'^2}{4}-\frac{B''}{2}+e^{2 A(z)} m_{5}^2$,
with $B= \left(d_0+ k^{\prime} c_0\right) \mathrm{ln}z$. 
It is found that for any $Dp-Dq$
system, $V(z)$ takes the general form of
\begin{eqnarray}
V(z)= \frac{1}{z^2}\left( \frac{(d_0+k^{\prime} c_0)^2}{4}+\frac{d_0+k^{\prime} c_0}{2}+m_5^2 \right).
\end{eqnarray}

In Table \ref{spectra-rho-DN}, we show the  meson and
glueball spectra by taking the boundary conditions as DN type,
$\psi_{n}|_{z=0} = 0\,,\,\,\,\, \partial_{z}
\psi_{n}|_{z=z_{m}} = 0\,$,
\textit{i.e.} the Dirichlet type at UV and Neumann type at IR.
It is found that in the $D3-Dq$ system, the predicted ${\bar q}q$ scalar meson is
below $1 {\rm GeV}$, and the scalar and tensor glueball
masses are much lighter than the lattice results. The predicted meson and glueball
masses are too light in the $D6-D4$ system and too heavy in the $D6-D6$ system,
and both cases are far away from experimental/lattice results.
The meson and glueball spectra in $D4-Dq$ brane systems are more
reasonable comparing with the experimental/lattice results. Especially the spectra of
scalar meson and scalar glueball in the $D4-D6$ and $D4-D8$ systems are very close
to the experimental/lattice results. The tensor glueball spectra in these two systems are
$80\%-90\%$ in agreement with the lattice results.

\begin{table}[th]
\begin{center}
$%
\begin{tabular}{|c|c|c|c|c|c|c|c|}
\hline
    & $Exp/Lat$ & $D_3-D_q$ &$ D_4-D_4$ & $ D_4-D_6$ & $ D_4-D_8$
& $ D_6-D_4$  & $ D_6-D_6$ \\
 \hline $z_m^{M}$ &  &$3.852$  & $2.04$ &$3.852$ &$5.268$ &$3.85281$&$1.453$\\
\hline $m_{\rho}$ &$0.77$ & $0.77^{*}$ & $0.77^{*}$ &  $0.77^{*}$ & $0.77^{*}$ &  $0.77^{*}$&$0.77^{*}$  \\
\hline $m_{f_0}$ & $1.37-1.5 $ &$0.893$  & $1.417$ & $1.584$  &$1.565$  &$0.548$  & $2.496$   \\
\hline $m_{G_0} $ & $1.6-1.7$ &$1.201$  &$1.956$  & $1.722$  &$1.633$  &$0.408$  &$2.858$    \\
\hline $m_{G_{2}} $& $\sim2.4$ & $1.920$ & $3.255$ & $2.255$  & $1.936$ &$1.442$ &$4.260$ \\
\hline $m_{G_{3}} $ & $\sim 3.69$ &$2.356$  &$4.240$  & $3.021$  &$2.684$  &$1.344$  &$6.131$    \\
 \hline
\end{tabular}%
\ $%
\end{center}
\caption{Results of the meson/glueball spectra in  the hard-wall $D_{p}-D_{q}$ system with the DN boundary condition. The unit for mass is in {\rm GeV}.}
\label{spectra-rho-DN}
\end{table}

In Table \ref{spectra-rho-DD}, we show the  meson and
glueball spectra by taking the boundary conditions as DD type,
$\psi_{n}|_{z=0} = 0\,,\,\,\,\,
\psi_{n}|_{z=z_m} = 0\,$,
\textit{i.e.} the Dirichlet type both at UV and at IR.
It is found that the results in hard-wall models are sensitive to the boundary conditions,
which is unlike the case in the soft-wall models as we have shown in Ref. \cite{HHYY}.
Using DD type boundary conditions, the predicted meson and glueball spectra in $D4-D8$ 
system is still close to the experimental/lattice results, but the error is bigger.  
We thus conclude that DN type boundary conditions are more appropriate for 
QCD hadron spectra.

\begin{table}[th]
\begin{center}
$%
\begin{tabular}{|c|c|c|c|c|c|c|c|}
\hline
    & $Exp/Lat$ & $D_3-D_q$ &$ D_4-D_4$ & $ D_4-D_6$ & $ D_4-D_8$
& $ D_6-D_4$  & $ D_6-D_6$ \\
\hline $z_m^{\rho}$ &  &$4.97624$  &$4.07999$&$4.97624$ &$5.8356$ &$4.976$ &$4.07999$ \\
\hline $m_{\rho}$ &$0.77$ & $0.77^{*}$ & $0.77^{*}$ &  $0.77^{*}$ & $0.77^{*}$ &  $0.77^{*}$&$0.77^{*}$  \\
\hline $m_{f_0}$ & $1.37-1.5 $ &$0.77$  & $0.939$ & $1.292$  &$1.441$  &$0.795$  & $1.744$   \\
\hline $m_{G_0} $  & $1.6-1.7$ &  $1.032$ &  $1.259$ & $1.404$   & $1.503$& $0.631$  &   $1.860$  \\
\hline $m_{G_{2}}$ & $\sim2.4$  & $1.637$ & $1.997$ & $1.837$  & $1.782$ &$1.532$  &$2.338$    \\
\hline $m_{G_{3}} $ & $\sim 3.69$ &$1.937$  &$2.441$  & $2.400$  &$2.444$  &$1.739$  &$3.726$    \\
\hline
\end{tabular}%
\ $%
\end{center}
\caption{Results of the meson/glueball spectra in  the hard-wall $D_{p}-D_{q}$ system with DD boudary condition. 
The unit for mass is in {\rm GeV}.}
\label{spectra-rho-DD}
\end{table}

\section{Summary}
\label{sec-summary}

We have investigated the light meson and glueball spectra in the $Dp-Dq$ hard-wall models, with
the IR cut-off fixed by the mass of vector meson mass $\rho$. We have used the  experimental/ lattice
results for the scalar meson mass in the range of $1370-1500$ and the scalar glueball mass in the 
range of $1600-1700$ as references.

We find that the $AdS_5$ hard-wall model, i.e. our $D3-Dq$ hard-wall model
is not the favored candidate of the \textit{holographic} QCD model, because the
predicted meson spectra and glueball spectra in this model does not agree well with
experimental/lattice results. The most favored candidates for the realistic \textit{holographic} QCD
model are the $D4-D6$ or $D4-D8$ hard wall models.  In these two models, the predicted meson
and glueball spectra are close to the experimental and lattice results. This picture
is in consistent with the curvature analysis in Sec. \ref{sec-DpDq}: For $D{3}$
background branes, the curvature is a constant, its dual
gauge theory is a conformal field theory, which is not QCD-like. 
The curvature for $D{4}$ background branes is small at IR, and large at UV, its dual 
gauge theory is strongly coupled at IR and weakly coupled at UV, which is similar to QCD. 

It is noticed that there is another scenario where the $\sigma(600)$ is identified as the scalar 
glueball  \cite{Scalar-lightGB}. This scenario can be realized in our $D6-D4$ system.
However, as we pointed out in Sec. \ref{sec-DpDq},  the curvature 
for $D{6}$ background branes is large at IR, and small at UV,
its dual gauge theory is weakly coupled at IR and strongly coupled at UV,
which is opposite to QCD.  Therefore,  the $D6-Dq$ system can be safely 
excluded for the candidates of \textit{holographic} QCD model. 

These results agree with the main findings in the $Dp-Dq$ soft-wall models 
\cite{HHYY}, where we find that  $Dp$ for $p=3,4$ systems are consistent with the 
Regge behavior of the vector and axial-vector mesons. More physical quantities need to
be evaluated and compared with experimental results in order to determine which  
$Dp-Dq$ system is more favored as the candidate of the realistic \textit{holographic} 
QCD model.

At the end, we want to point out that our results are based on the assumption that the 5D
mass square of the dual field follows the relation Eq. (\ref{Eq-m5}) in the AdS/CFT dictionary. This relation
might be modified in the non-conformal $Dp-Dq$ systems. We need further  studies
along this direction.

\vskip3mm \textit{\textbf{Acknowledgments: ---}} We thank valuable discussions with
Y. Chen, Q. Zhao and B.S. Zou on scalar mesons and glueballs. The work of M.H. is supported by
CAS program "Outstanding young scientists abroad brought-in", the CAS key
project under grant No. KJCX3-SYW-N2, and NSFC under grant No. 10875134 and
No. 10735040. Q.S.Y. thanks the hospitality of TPCSF during his stay when this work was
initiated. The work of Y.Y. is supported by National Science Council (NCS) of Taiwan
(97-2112-M-009-019-MY3) and National Center for Theoretical Sciences (NCTS) through
NCS of Taiwan.

\end{document}